\documentclass[useAMS]{mn2e}
\usepackage{graphicx}
\usepackage{epstopdf}
\usepackage{color}

\title[sSFR and the metallicity]{On the relation between sSFR and metallicity}

\author[Pipino et al.]{A. Pipino\thanks{antonio.pipino@phys.ethz.ch}$^1$, S.J. Lilly$^1$, \& C.M. Carollo$^1$\\
$^1$Institute for Astronomy, ETH Zurich, Wolfgang-Pauli-Strasse
  27, 8093 Zurich, Switzerland}

\long\def\symbolfootnote[#1]#2{\begingroup%
\def\thefootnote{\fnsymbol{footnote}}\footnote[#1]{#2}\endgroup}

\begin{document}

\date{Accepted 2014 March 20.  Received 2014 March 18; in original form 2014 January 10}

\pagerange{\pageref{firstpage}--\pageref{lastpage}} \pubyear{2008}

\maketitle

\label{firstpage}

\begin{abstract}
In this paper we present an exact general analytic expression $Z(sSFR)={y_Z \over \Lambda(sSFR)}+I(sSFR)$ linking the gas metallicity $Z$ to the specific star formation rate (sSFR),
that validates and extends the approximate relation put forward by Lilly et al. (2013, L13), where $y_z$ is the yield per stellar generation,
$\Lambda(sSFR)$ is the instantaneous ratio between inflow and star formation rate expressed as a function
of the sSFR, and $I$ is the integral of the past enrichment history, respectively.

We then demonstrate that the
instantaneous metallicity of a self-regulating system, such that its sSFR decreases with decreasing redshift, can 
be well approximated by the first term on the right-hand side in the above formula, which provide an upper bound to the metallicity. 
The metallicity is well approximated also by $Z_{L13}^{id}=Z(sSFR)={y_Z \over 1+\eta+sSFR/\nu}$ (L13 ideal regulator case), which provides a lower bound to the actual metallicity.
We compare these approximate analytic formulae to numerical results and infer a discrepancy $<0.1$ dex in a  range of metallicities ($log (Z/Z_{\odot})\in [-1.5,0]$, for $y_z\equiv Z_{\odot}=0.02$) and almost three orders of magnitude in the sSFR. 

We explore the consequences of the L13 model on the mass-weighted metallicity in the stellar component of the
galaxies. We find that the stellar average metallicity lags $\sim 0.1-0.2$ dex behind the gas-phase metallicity relation, in agreement with the
data.

\end{abstract}


\section{Introduction}
The evolution of the metallicity in galaxies constrains the
history of the gas accretion relative to the star formation, as well as the relative importance of outflows. As
such it has been extensively studied at different cosmic epochs.
Whilst the full stellar metallicity distribution is available only for a few selected nearby galaxies, including the Milky Way and its components,
average metallicities in the stars and in the gas of star forming regions are available for many more objects
at different cosmic epochs (O'Connell, 1976, Lequeux et al., 1979, Tremonti et al., 2004, Savaglio et al. 2005, Mayer et al., 2005, 2006, Erb et al. 2006, Cid Fernandes et al. 2007, Maiolino et al., 2008, Mannucci et al., 2009, Zahid, et al., 2012, Leja, et al., 2013, Gallazzi et al., 2006, Panter et al., 2008, Sommariva et al., 2010, and references therein). These observations have established that, at any redshift $z<4$, the most massive galaxies are the most metal rich in both their gas and stellar components.
Moreover, at fixed mass, the gas metallicity of star forming objects decreases with increasing redshift (Erb et al., 2006, Maiolino et al., 2008,
Mannucci et al., 2009, Mannucci et al., 2011 Richard et al., 2011, Yuan et al., 2013).

Among the many theoretical attempts to understand the drivers of such relations, 
analytical chemical evolution models appeal either to a decreasing importance of outflows (e.g. Garnett, 2002, Tremonti et al., 2004,
Spitoni et al., 2010, and/or { differential winds}\footnote{Namely the outflows in which the ejection of some chemical
species is enhanced with respect to others} , e.g., Dalcanton, 2007, Recchi et al., 2008) or to an increase in the star formation efficiency 
(Dalcanton, 2007,
Spitoni et all., 2010, Peeples \& Shankar, 2010),
a variation in the yield (via a flattening of the IMF, e.g. Koeppen et al., 2007) or an increase
in the fraction of re-accreted metals (Dav\'e et al., 2012), with galactic mass, as possible explanations.
It is worth pointing out that, in many cases, these models are adopted to interpret the data at a single
epoch.
On the other hand, galaxy formation numerical experiments, such as cosmological simulations 
and semi-analytical models, despite qualitatively matching 
the $z\sim 0$ relation, do not reproduce its slope (e.g. Pipino et al., 2009) and generally suffer from over-predicting the metallicity
of high-redshift star forming galaxies (see, e.g., the discussion in Maiolino et al., 2008, and references therein, Sakstein et al., 2011, Yates et al., 2012).

{ More recently, a new dimension was added to the observational picture, with studies suggesting
that, \emph{globally}, the gas metallicity $Z$ of $z\sim 0$ galaxies depends
also on the (specific) star formation rate (SFR): at a fixed galaxy mass, higher metallicities correspond
to a lower star formation activity (e.g. Ellison
et al., 2008, Mannucci et al., 2010, but see e.g. Yates et al., 2012). Furthermore, there
is empirical evidence suggesting the \emph{local} nature of such mass-SFR-Z relation (Rosales-Ortega et al., 2012), and
the question becomes as to whether the $Z=Z(M,SFR)$ relation is redshift independent; that is if 
high redshift galaxies populate the extrapolation of the so-called $z\sim0$ fundamental-metallicity relation,
that can either be a surface in the mass-SFR-metallicity space 
(Mannucci et al., 2010) or a plane (Lara-Lopez et al., 2010, 2013) out to $z\sim2$, 
or even $z=3$ (when accounting for changes in the ionization parameter, Nagajima \& Ouchi, 2013, Cullen et al., 2013). 
Such a debate is lively and far from being set (see e.g., Cresci et al., 2011, Richard, et al. 2011, Niino, 2012, Wuyts et al., 2012, Christensen et al., 2012, Stott et al., 2013, Henry et al., 2013, Troncoso et al., 2013, Zahid et al., 2013, Belli et al., 2013). However, irrespective of the final answer, it highlights the importance
of fully and simultaneously addressing galaxy evolution in terms of mass-metallicity and mass-SFR relations and their
evolution with redshift. 
It offers an independent test-bed to the above-mentioned analytic chemical evolution models and provides 
new constraints to the the increasing body 
of theoretical works (e.g. Bouch\'e et al., 2010, Dutton et al., 2010, 
Lilly et al. 2013, L13, and references therein) aimed at explaining the existence and the evolution of the SFR-mass relation
(e.g., Elbaz et al., 2007, Noeske et al., 2007, Daddi
et al., 2007, Pannella et al., 2009, Oliver et al., 2010) and/or the cosmic run of the specific SFR (sSFR) with simple models for the galaxy growth.

In particular, L13 show that the $z<2$ evolution of the sSFR of galaxies may be
controlled by the cosmological infall of gas, through the regulating action of the gas reservoir via
a Schmidt (1959) linear star formation law. Such a simple model broadly explains at the same time the cosmic evolution
of the sSFR (e.g. Gonzalez et al., 2010, Stark et al., 2012, and references therein) and the stellar-to-halo mass ratio (e.g., Moster et al., 2010).
More importantly, the L13 model has the additional appealing property of offering
an explanation both to the evolution of the gas phase metallicity and to its scatter
at a given epoch by directly linking it
to variations in the sSFR with just one equation. That is, the sSFR both enters as a second paramenter
in setting the metallicity and gives and explanation to the epoch-invariant fundamental metallicity
relation, thereby linking the epoch-dependence and the SFR-dependence of the mass-metallicity relation.

At a fixed epoch, the slope of the  mass-metallicity relation is then given by the variation of
both the star formation and outflow efficiencies with stellar mass (see also Calura et al., 2009).
In L13, however,
{ the instantaneous gas phase metallicity is replaced with the value derived considering the system in equilibrium
(i.e. imposing $dZ/dt=0$) for both an ideal case of regulator (steady state at constant gas fraction)
and a case in which the gas fraction is slowly changing}.\\
\indent Other analytic models do not either make explicit the sSFR dependence of metallicity (e.g. Dayal et al., 2013) or, despite their similarity to L13, adopt a different notion
of ``steady-state'' (e.g., constant gas mass evolution, e.g., Dav\'e et al., 2012), claiming that the temporal variation in the metallicity
for a given galaxy is driven by the amount of metals ejected in the surrounding medium and then re-accreted.
Also in the case of Dav\'e et al. (2012), approximate values for the metallicity are adopted.\\
\indent Given the important role of metallicity as a constraint to galaxy formation theories and
the progresses in the measurement of Z, SFR and stellar masses at progressively higher redshifts, it
is important to derive full analytical expressions
that link the metallicity evolution to the sSFR evolution of a single galaxy for generic
gas accretion and outflow histories. If correct, the above mentioned approximated formulae (e.g. L13, Dav\'e et al., 2012)
could be then re-derived from such general solutions as special cases and applied in suitable regimes
of the galaxy growth.

\begin{table*}
\centering
\begin{minipage}{120mm}
%\scriptsize
\begin{flushleft}
\caption[]{Input parameters}
\begin{tabular}{c | c | c | c}
\hline
\hline
		& This paper 		& L13     & remarks\\
\hline
Baryonic accretion rate  &   $\dot{M}_{acc}$     &  $\Phi$ & given by cosmological background \\

SF efficiency   &   $\nu$    		& $\epsilon$  & constant (may vary with galactic mass and/or cosmic time) \\

Gas-to-total fraction  &  $\mu$         & $\nu_{gas}$   & - \\

Gas-to-star fraction  &  $f$           & $\mu$   & - \\

Infall rate-to-SFR ratio & $\Lambda$   &  $1/f_{star}$ & { varies with time} \\

Outflow rate-to-SFR ratio & $\eta$   &  $\lambda/(1-R)$ & { may vary with time} \\

Yield per stellar generation &   $y_z$  &   $y$   & constant\\

Metallicity of infalling gas &  $Z_A$   &  $Z_0$  & constant \\

Returned fraction &    $R$ &    $R$ & constant \\

\hline
\end{tabular}
\end{flushleft}
\end{minipage}

 \label{t1}

\end{table*}
The aim of this paper is thus to validate and extend the L13 relation between $Z$ and sSFR. 
To this end, we revisit the L13 equations, link them to analytical models of
chemical evolution  (Pagel \& Patchett 1975; Hartwick 1976; Twarog 1980; Tinsley, 1980, Matteucci \& Chiosi, 1983,
Clayton, 1988, Edmunds, 1990, Koeppen \& Edmunds, 1999, Matteucci 2001, Spitoni et al., 2010),
and derive a more general relation 
in which the metallicity Z is an explicit function of the sSFR for arbitrary gas inflow
and outflow histories. We then derive simplified relation for the
the Closed Box Model, the steady state evolution and the L13 model as special cases of
the general solution. 
Furthermore, we study the range
of validity and the goodness of the L13 approximation by
comparing these results to a direct numerical integration of the
same equations as well as to the predictions of full numerical chemical evolution
models, which relax some of the assumptions
done to make the problem analytically tractable.\\

\indent Finally, we test the predictions of the L13 model for the evolution of the mass-stellar metallicity relation 
with redshift in the specific L13 case and compare it to recent observations (Sommariva et al., 2012).

The L13 model and some of its equations are briefly summarized in Sec. 2 with the double aim to both set the stage,
introducing the relevant physical quantities and symbols, and to link it to the standard
equation of analytic chemical evolution model. In Sec. 3, 
general relations between gas-phase metallicity and sSFR are presented and their special cases
discussed. L13 model predictions regarding the stellar average metallicity and
its comparison to data are presented in Sec. 4. Finally,
in Sec. 5 we summarize and discuss our main conclusions. 

\section{The L13 model in the framework of analytic chemical evolution}

\subsection{The regulation of the baryonic content in L13}

L13 suggests that the average galaxy evolution can be broadly explained by very simple physics, as it is determined by the host halo
accretion rate and regulated by the star formation
rate $\psi$, which is directly proportional to the interstellar medium mass (the ``gas reservoir'') via the star formation efficiency
$\nu$ (see Table~\ref{t1}), that we will
keep constant with time.
The model also accounts for the action of SFR-related (e.g. supernova-driven) galactic winds.

Let us define $\mu$ as the gas fraction $M_{gas}/M_{tot}$, with $M_{tot}=M_* +
M_{gas}$.
This implies that the sSFR can be written as (Eq. 7 in L13)\footnote{See the Appendix for a more general
version when a Schmidt law with exponent $1+x$ is adopted}:
\begin{equation}
sSFR=\psi /M_* = \nu \, M_{gas}/ M_* = \nu {\mu \over 1-\mu}\, .
\label{eq:ssfr1}
\end{equation}

The system accretes gas via inflows, with a given accretion rate $\dot M_{acc}$. %\footnote{$\Phi$ in L13 formulation}.
More specifically, the L13 model assumes that haloes grow according to average prescriptions given by fit to numerical 
simulations (e.g. Eq 3 in L13 and references therein). Baryons are accreted proportionally
to the dark matter, via the universal baryon fraction. A fraction $f_{gal}$ of the accreted baryons 
can penetrate into the actively star forming region and be transformed into stars, as well as
possibly ejected by winds.

{ The simplest (ideal) case of such a regulator has the feature of setting the sSFR equal to the specific baryonic
accretion rate $sMIR_B$}. L13 (c.f. their Fig. 3) further show that, for any sudden and instantaneous variation in
$sMIR_B$, the sSFR adjusts to a value that coincides with $sMIR_B$ on a timescale set by the shorter between $1/\nu$
and $1/sSFR$.
The sSFR tracks the $sMIR_B$ also if this is steadily increasing with time. Only when the variation 
$d sMIR_B /d t < 0$ and occurs on a timescale which is faster than $1/\nu$, then the sSFR decrease is slower
than the drop in the $sMIR_B$.

Our study will focus on the metallicity-sSFR dependence. Therefore we do not further discuss the dark
matter host halo growth. The cycle of inflow-star formation-outflow that we discuss below
pertains to the baryons within the galaxy, therefore our conclusions do not depend on the chosen
$f_{gal}$ either, with the assumption that this value is not affected by, e.g., the galactic star formation rate or the outflows.

Below we briefly present the relevant equations of L13's model set-up in a slightly different way in order to link the L13 equation
and symbols to the terms that are more common in chemical evolution studies and with the aim
of summarizing some of L13's key results to the reader, setting the stage for the present study.
To this purpose, in
Table~\ref{t1} we summarize the main physical quantities and the symbols adopted in both the present work and in L13.

\subsection{Basic equations for the evolution of gas mass and metallicity}

As in L13, in this paper we follow the evolution of a galaxy made of gas, assumed
to be in a single phase and well mixed at any time, with initial mass $M_{gas,0}$, and stars, whose initial
mass is set to zero. 
The evolution of the system can be studied solving an array of equations representing
the conservation of the total, the gas and the metal mass in presence of source terms (infall, outflow and nucleosynthesis).

Before doing it, it is convenient to
introduce the variables $\Lambda(t)$ and $\eta(t)$
%\footnote{These quantities are equivalent to $1/{f_{star}}$ and $\lambda/(1-R)$ in L13's formulation, respectively. As for $\nu$, we chose to adopt the symbols more frequently adopted in the chemical evolution}
, defined in order for the infall
and outflow rate to be cast in terms of the SFR $\psi (t)$.
Namely, the outflow rate $W (t)$
is defined as (Matteucci \& Chiosi, 1983):
\begin{equation}
W (t) = \eta(t) (1 - R) \psi (t),
\label{eq:w}
\end{equation}
\noindent
and it is justified by the observational evidences of ubiquitous winds in 
star forming galaxies (e.g., Newman et al., 2012, Bordoloi et al., 2011, 2013, Weiner et al., 2005), with mass loading factors
comparable to the SFR. { Since the same loading factor, within the uncertainties, is observed at different redshifts in galaxies with different SFRs (c.f. Newman et al., 2012 and references therein),  for a first order approximation it is 
reasonable to assume that $\eta\sim const$ in galaxies with SFR-driven winds. Therefore, 
in the following we will present both examples and special cases assuming that $\eta$ is strictly constant in time. On the other
hand, in deriving the full solution $Z=Z(sSFR)$ we will let $\eta$ arbitrarily vary with either time or sSFR.}

The infall rate is given by:
\begin{equation}
\dot M_{acc}  = \Lambda(t) (1 - R) \psi (t).
\label{eq:a}
\end{equation} 
\noindent
The term $\Lambda(t)$ was introduced and set to a constant value by Matteucci \& Chiosi (1983), 
to make the problem tractable analytically. { The same time-invariant definition (i.e. Eq.~\ref{eq:a} with $\Lambda=const$) is adopted in many
other papers in the literature. In our approach, instead, the equation above is actually inverted and solved for $\Lambda(t)$, which
will provide a way for parameterize
how the SFR \emph{instantaneously} responds to changes in the known accretion rate. Therefore, in this case (and in L13's formulation) $\Lambda(t)\propto {\dot M_{acc} \over \psi (t)}$ }is a function
of time and becomes the instantaneous measure of the ratio between the gas accretion rate (given by
the cosmological model) and the star formation rate.

The returned fraction by stars $R$ is defined by 
invoking the Instantaneous Recycling Approximation (IRA, Schmidt, 1963) as:
\begin{equation}
R = \int_{1\, M_{\odot}}^\infty (m - m_R) \phi (m) dm,
\label{eq:r}
\end{equation}
\noindent
where $\phi (m)$ is the IMF and $m_R$ is the mass of the stellar remnant.
The IMF is assumed constant in time. 
The yield per stellar generation is then defined as (Tinsley, 1980):
\begin{equation}
y_Z = {1 \over {1 - R}} \int_{1\, M_{\odot}}^\infty m p_{Z, m} \phi (m) dm,
\label{eq:yield}
\end{equation}
\noindent
where $p_{Z, m}$ is the fraction of newly produced and ejected metals
by a star of mass $m$. 
Finally, $Z_A$ is the
metallicity of the infalling gas.

Under these assumptions and definitions, the equations that regulate the evolution of the system in L13 become
exactly those used by analytical models for chemical evolution with inflows
and outflows (e.g., Pagel \& Patchett 1975, 
Hartwick 1976, Twarog 1980; Tinsley, 1980, Edmunds, 1990, Koeppen \& Edmunds, 1999, Matteucci 2001, Spitoni et al.,
2010):

\begin{equation}
\cases{{d M_{tot} \over d t} = (\Lambda(t) - \eta(t)) (1 - R) \psi (t) \cr
\cr
{d M_{gas} \over d t} = (\Lambda(t) - \eta(t) - 1) (1 - R)\psi (t) \cr
\cr
{d Z \cdot M_{gas} \over d t} = (1 - R) \psi (t) [\Lambda(t) Z_A + y_Z - (\eta(t) + 1) Z]}
\label{eq:system}
\end{equation}
\noindent with the only difference that, in L13 and in this paper, $\Lambda$ and $\eta$ may vary with time,
not least as the system increases its mass.
Also, both the star formation law and the accretion rate of the galaxy are specified, whereas in standard analytical chemical evolution models, it is customary
to express metallicity variations as a function of $\mu$ without any explicit dependence on both SFR and $\dot M_{acc}$.

Combining the first two equations in the array~\ref{eq:system}, L13 (see their Eq. 12) came to the following link between $\Lambda(t)$ and sSFR:
\begin{equation}
\Lambda(t)= sSFR(t)/\nu +\eta(t)+1+{1 \over (1-R)\nu} {d ln f \over dt} \, ,
\label{eq:link}
\end{equation}
where $f=M_{gas}/M_*=sSFR/\nu$.
Eq.~\ref{eq:ssfrsteady} that we discuss below is a special case of this general relation.

\subsection{The gas metallicity in the L13 approach}

Rather than explicitly solving  Eq.~\ref{eq:system} also for the metallicity, L13 derive approximate
solutions by assuming the system to be in equilibrium. { That is by solving the third equation
in the array~\ref{eq:system} by imposing $dZ/dt=0$.
Indeed, L13 derive two slightly different approximations for the metallicity: 
one that holds in the case of a ideal regulator (i.e. with the gas fraction
identically constant in time), and one that holds for the more realistic non-ideal regulator (Eqs.~26 and~29 in L13).
In the formalism of this paper, these two approximations become:\\
\begin{eqnarray}
Z_{L13}^{id}\equiv {Z_A + y_z \over sSFR/\nu +\eta+1} \cr \cr \cr
Z_{L13}\equiv {Z_A + y_z \over sSFR/\nu +\eta+1+{1 \over (1-R)\nu} {d ln f \over dt} }={Z_A + y_z \over \Lambda(t)}\, ,
\label{appr2}
\end{eqnarray}
}
\noindent respectively. We remind the reader that in L13 $\eta=const$.
Such simple expressions for the metallicity have non-trivial consequences.
In the first place, the variation in the sSFR with cosmic epoch will drive
a change in the metallicity of a given ``average'' galaxy.

Secondly, at any given time, two galaxies with the same stellar mass may have different metallicities,
according to the values of the $sSFR,\, \eta ,\, \nu,\, f$ that characterize them.
In the framework of this analytical model, such a difference in metallicity 
is caused by the different ``equilibrium'' gas fraction in the two galaxies (or equivalently, their sSFR, if the
other terms in the denominator are smaller). 
That is, a mass-metallicity-SFR relation 
is naturally predicted by the L13 model.
{ More quantitatively, at a given epoch, two galaxies $i=1,2$ with given stellar masses $M_i$ 
will have the ratio of their metallicities given by (assuming $Z_A=0,\eta=0$, ideal regulator case):
\begin{eqnarray*}
%\cases{
Z_{L13,1}/Z_{L13,2} = \cr
\cr
= \nu_1 M_1/\nu_2 M_2 \;(SFR_2+M_2\cdot \nu_2) / (SFR_1+M_1\cdot \nu_1) = \cr
\cr
=: M_1/M_2 \; SFR_2/SFR_1 \,H(sSFR_i,\nu_i(M_i)) \, , %}
\label{fund1}
\end{eqnarray*}
that is
$$Log (Z_{1}/Z_{2}) = Log (M_1/M_2) - Log (SFR_1/SFR_2) + Log (H)\, .$$}

%~~~~~~~~~~~~~~~~~~~~~~~~~~~~~~~~~~~~~~~~~~~~~~~~~~~~~~~~~~~~~~~~~~~~~~~~~~~~~~
\section{The metallicity evolution}

The set of equations presented in (\ref{eq:system}) with suitable initial conditions are sufficient to characterize
the galaxy evolution in terms of its gas mass, gas fraction and metallicity evolution, by direct integration
over time. 
These solutions can then be transformed into an sSFR dependence via the sSFR-$\mu$ (Eq.~\ref{eq:ssfr1}) and sSFR-$\Lambda$ (Eq.~\ref{eq:link})
relations. This will be the core and novel aspect of this paper. In particular, we will derive two versions of a more general solution of the equations which include among their terms both the L13 approximations ($Z_{L13}^{id}$, $Z_{L13}$). We will then show under which
conditions and how quickly the other terms
is the solutions become negligible and, thus, L13 approximations become good solutions. 
In the final part of this Section we will compare the full analytical solution
to the approximate L13 solution, and both to numerically-derived trends, to test the range of validity and the accuracy.

\subsection{General solution: $Z$ as a function of time and sSFR}

Let us take a step back and start by recalling the  
formal general solutions for the gas and the metallicity evolution 
with the explicit time dependence that can be derived by the
same set of three equations (6) if one leaves the accretion rate unspecified and further assumes
$\eta=const$ (in analogy with previous works and L13):
\begin{eqnarray}
M_{gas}=\displaystyle e^{-\nu (1-R)(1+\eta) t} \times \cr
\times \Bigg(M_{gas,0}+\int_0^t \! e^{\nu (1-R)(1+\eta) t'} \dot Macc(t')\mathrm{d} t'\Bigg)\, ,
\end{eqnarray}
\noindent and
\begin{eqnarray}
Z=\displaystyle y_z \nu (1-R) e^{-\nu (1-R)\,\int_0^t \! \Lambda(t') \mathrm{d} t'} \times \cr
\times \int_0^t \! e^{\nu (1-R)\,\int_0^{t'} \! \Lambda(t'') \mathrm{d} t''} (1+{\Lambda(t') Z_A(t')\over y_z}) \mathrm{d} t' \, ,
\label{eq:sol}
\end{eqnarray}
\noindent (see also Recchi et al., 2008, for the solution with $Z_A=0$).
For simplicity, we assumed that $Z(0)=0$ and that $\nu$ and $\eta$ are constant with time.
We also assumed no variations in the IMF (and hence in the yield $y_z$) with either time or mass. 
These formulae can be obtained as standard solutions of the differential equations of the array~\ref{eq:system}
in a manner that is similar to what we show in Section 3.3 (Eq. 19 onward); therefore we do not repeat
the derivation here.

Assuming that $Z_A=const$, 
substituting $\Lambda(t)$ in Eq.~\ref{eq:sol} with the expression given by Eq.~\ref{eq:link} and integrating the resulting
expression by parts, it follows that:
\begin{eqnarray*}
%Z=\displaystyle y_z \nu (1-R) 
Z=Z_A + y_z \nu (1-R) 
\Bigg( {1 \over \nu (1-R) (1+\eta+sSFR/\nu)} + \\
- {e^{-\nu (1-R)\,((1+\eta)t+\int_0^t \! sSFR(t')/\nu \mathrm{d} t')} \over \nu (1-R) (1+\eta+sSFR(0)/\nu)} sSFR(0)/\nu + \\
\\
- {(1+\eta)\, e^{-\nu (1-R)\,((1+\eta)t+\int_0^t \! sSFR(t')/\nu \mathrm{d} t')} \over sSFR/\nu } \times \\
\times \int_0^t \! {e^{\nu (1-R)\,((1+\eta)t'+\int_0^{t'} \! sSFR(t'')/\nu \mathrm{d} t'')} \over \nu (1-R) (1+\eta+sSFR/\nu)^2}{\mathrm{d} sSFR/\nu \over \mathrm{d} t'} \mathrm{d} t' \Bigg)= \\ 
\\
=: (I_1-I_2-I_3) \, .
\label{sol1}
\end{eqnarray*}
\noindent \noindent Where
\begin{equation}
I_1\equiv Z_{L13}^{id}={Z_A+y_z \over 1+\eta+sSFR/\nu} \, . 
\label{appr1}
\end{equation}
\noindent  Despite two terms in the addition ($I_2 , I_3$ - which incorporate the integral of the accretion history), 
have still the explicit dependence on time, we made an important step forward as we have the first 
term ($Z_{L13}^{id}$) depending only on the sSFR. As we will see below, $Z_{L13}^{id}$ is also a bounding value to the true
metallicity. To understand the meaning of $I_1=Z_{L13}^{id}$ in this context, we need to look first at the following special case of the general solution: the evolution
at constant gas fraction.

\subsection{Special case I - Evolution at constant gas fraction}

If the galaxy is constantly in an accretion-dominated regime, that is
if we add the assumption that $\Lambda,\, \eta$ are both constant with time and that $\Lambda-\eta>$1, 
then $M_{tot}$ increases with time.
The gas fraction evolves as (e.g., Eq. 9 in Recchi et al., 2008):
\begin{equation}
\mu= \mu_{\rm steady} + {f_{M}\over \Lambda - \eta}\, ,
\end{equation} 
where
\begin{equation}
\mu_{\rm steady} = {{\Lambda - \eta - 1} \over {\Lambda - \eta}} \, ,
\label{musteady}
\end{equation}
and
\begin{equation}
f_{M}=M_{gas,0}/M_{tot}(t)\, .
\end{equation}
Therefore, the gas fraction tends to the value given by $\mu_{\rm steady}$.
This is what we refer to as \emph{steady state}, namely an evolution at a constant gas fraction,
whose value is set by the constant $\Lambda-\eta$.
In this particular case it is trivial to combine Eqs.~\ref{eq:ssfr1} and~\ref{musteady} to show
that the sSFR can be expressed as:
\begin{equation}
sSFR/\nu=\Lambda-\eta-1 \, .
\label{eq:ssfrsteady}
\end{equation}
We note that the convergence towards the asymptotic values is faster for larger $\nu$ (the shorter the star formation timescale) and/or larger $\Lambda$.

When the steady state is attained, with $\Lambda, \eta$ and the $sSFR$ constant in time,
the solution for the metallicity becomes much simpler:
\begin{eqnarray}
Z_{ss}^{true}=Z_A + \frac{y_z }{\Lambda} \Bigg(1-e^{-\nu (1-R)\Lambda t}\Bigg)\, \cr 
\rightarrow Z_A+\frac{y_z }{\Lambda}=Z_A+\frac{y_z }{1+\eta+sSFR/\nu} \, ,
\end{eqnarray}
at times $t> 1/ \nu > 1/(\nu (1-R) \Lambda)$.

That is, the metallicity settles to the constant value given by $Z_{L13}^{id}$.
As such, the asymptotic regime for the equilibrium metallicity (i.e. when $dZ/dt=0$) is used in L13 as the value
for the metallicity in the case of the ideal regulator (i.e. when the gas fraction stays constant).
This result further clarifies the meaning to the $Z_{L13}^{id}$ term (Eq. 11) contributing to the metallicity in the general equation.
It is in fact a ``steady-state-like'' term, determined by the current value of the sSFR.

We also derive another interesting result, probably overlooked in the recent literature
on the sSFR evolution at high redshift. In particular, Eq. 15 implies that one can easily model
a galaxy evolving at constant sSFR as the result of an accretion
dominated regime where $\Lambda-\eta-1 = const > 1$.
The results showed in this section imply that,
at the same time, the metallicity would not evolve (assuming $Z_A=const$).
Since $Z$ is observed to decrease at $z>2$, this is another reason to suspect that the sSFR 
is also not constant at $z>2$ (see, e.g., Stark et al., 2012).

\subsection{Integrating the metallicity equation over the sSFR}

One can alternatively
set up the differential equation for the metallicity variation as a function
of the sSFR as the time variable.
{ With the aim to derive a very general solution, from this section onward, not only we keep considering $\Lambda$ as a function
of time (and sSFR), but also we relax the assumption of $\eta=const$ with both time and sSFR.}
In particular, we note
that Eq.~\ref{eq:link} can be rewritten as:
\begin{equation}
{d sSFR \over dt}=\nu (1-R) sSFR (\Lambda(t)-\eta(t)-1-sSFR/\nu) \, .
\label{eq:denom}
\end{equation}
Combining the third equation with the second one in the array~\ref{eq:system},
one can write:
\begin{equation}
{d Z \over d t} = (1 - R) \nu [\Lambda(t) Z_A + y_Z - \Lambda(t) Z]\, .
\label{eq:z}
\end{equation}
\noindent Dividing Eq.~\ref{eq:z}
by Eq.~\ref{eq:denom}, one can derive an expression for $dZ/dsSFR$.
Some algebra then leads to the following differential equation for the metallicity:
\begin{equation}
{dZ \over dsSFR} + F(sSFR) Z= y_z G(sSFR) \, ,
\end{equation}
where
\begin{equation}
G(sSFR)={1+{Z_A \Lambda \over y_z} \over sSFR/\nu (\Lambda-\eta-1-sSFR/\nu)} \, ,
\end{equation}
and
\begin{equation}
F(sSFR)={1\over{{1\over \Lambda}+{Z_A \over y_z}}}\,G(sSFR) \, .
\end{equation}
This equation has the following formal solution:
\begin{eqnarray}
Z=\displaystyle y_z e^{-\,\int_{x_0}^x \! F(x') \mathrm{d} x'} \int_{x_0}^x \! e^{\int_{x_0}^{x'} \! F(x'') \mathrm{d} x''} G(x') \mathrm{d} x' \, ,
\end{eqnarray}
where $x=sSFR/\nu$.
Integrating by parts, with the further assumption that $Z_A=const$, the solution can be written as:
\begin{equation}
Z=Z_A+{y_z \over \Lambda(x)} - y_z e^{-\int_{x_0}^x F(x')dx'}
\int_{x_0}^x {e^{\int_{x_0}^{x''} F(x')dx'} \over \Lambda(x'')^2} {d\Lambda \over d x''} dx''  \, .
\label{appendix}
\end{equation}

This new way to solve for the metallicity uses the sSFR itself as the time
variable. This formal analytic solution is similar to Eq.~\ref{sol1}\footnote{The formal derivations of
the solutions are exactly the same.}, with the difference
that here we make explicit 
the contribution by the instantaneous value of $\Lambda(t)$ and $\eta(t)$
as if it were in the steady state, that is 
\begin{eqnarray}
Z_{ss}^{inst}=Z_A + {y_z \over \Lambda}=\cr \cr
=Z_A + {y_z \over sSFR/\nu +\eta+1+{1 \over (1-R)\nu} {d ln f \over dt} }\equiv Z_{L13}\, ,
\label{appr2}
\end{eqnarray}
\noindent 
namely the value of the metallicity adopted in L13 for the ``non-ideal regulator'' (i.e.
gas fraction slowly varying in time).
The other term in the addition is the 'resistance' to move to the new steady state, $- \tilde I_2 $, 
given by the past chemical evolution history.
With this version of the general solution for the metallicity, we made
explicit the fact that $Z_{L13}$ is one of the
terms that contribute to the actual metallicity of the galaxy in the general case. 
Eq.~23 readily tells us that, when $\tilde I_2$ is small (as in the L13 model), the evolution of galaxies can be
approximated by a sequence of steady state solutions with lower equilibrium gas
fractions (lower sSFR) and higher metallicities. We quantify this statement in the next section.

\subsection{Special case II - Accretion rate slowly changing with time (L13)}

Having derived general formulae linking the metallicity to the sSFR for arbitrary gas accretion
histories (Eqs.~11 and~23), we can now discuss the L13 approximations.
In order to move from the general equations discussed above to the L13 special cases
we simply need to add the assumption - explicitly made in L13 - that the sSFR slowly changes in time
in order to better quantify the other terms ($I_2$, $I_3$, $\tilde I_2$) in both Eqs.~11 and~23.
In particular, we now discuss the case in which the sSFR slowly decreases following the cosmological
decrease in $sMIR_B$.
As a matter of fact, similar considerations can be done for a slowly increasing sSFR, driven by,
e.g., an increasing accretion rate; therefore we do not further discuss this specific case and
refer the reader to L13 (c.f. their Fig. 3) for some examples on how quickly the sSFR responds 
to changes in either direction in the accretion rate. For simplicity, we also assume that $Z_A=0$, since
it only adds a constant offset.\\

\noindent For a smoothly declining accretion history it follows that:

\noindent 1) { As discussed in L13 (their Eq. 39) ${d ln f \over dt}$ is small, but finite and negative}, hence:
\begin{equation}
\; {1 \over sSFR/\nu +\eta+1+{1 \over (1-R)\nu} {d ln f \over dt} }> {1 \over sSFR/\nu +\eta+1}
%Z={y_z \over \Lambda(t)}={y_z \over sSFR/\nu +\eta+1+{1 \over (1-R)\nu} {d ln f \over dt} }> {y_z \over sSFR/\nu +\eta+1}\, .
\end{equation}
Therefore $Z_{L13}^{id}<Z_{L13}$. Moreover:

\noindent 2) $\tilde I_2 >0$. Therefore $Z<Z_{L13}$.\\

\noindent 3) Also, $I_2$ falls off exponentially. Therefore $I_2 \simeq 0$,
for practical purposes, whereas $-I_3 >0$, as the time derivative of the sSFR is negative. Therefore
$Z>Z_{L13}^{id}$\\

\noindent By combining these results together, we derive that the true value of the metallicity is always within the range
$[Z_{L13}^{id},Z_{L13}]$.
%:=[{{y_z \over sSFR/\nu +\eta+1}, {y_z \over sSFR/\nu +\eta+1+{1 \over (1-R)\nu} {d ln f \over dt} }}] $$
This shows that, for a large class of models with the sSFR slowly decreasing in time, rather simple expressions (Eqs.~\ref{appr1}-~\ref{appr2})
can be used to bracket the actual gas-phase metallicity.
In other words, for $\dot M_{acc}$ decreasing with time, the true metallicity will be bound between $Z_{L13}^{id}$ (the steady-state-like
metallicity set by the current value of the sSFR, the ideal regulator case in L13) and $Z_{L13}$ (the steady-state metallicity
set by the current value of $\Lambda(t)$, in L13 this is the formula used when the gas fraction is allowed to vary in the non
ideal regulator case).

The next step is to assess
the goodness of the approximation of using the steady-state(-like) metallicities (i.e. either $Z_{L13}^{id}$ or $Z_{L13}$)
as an estimate of the current metallicity for these systems. This will be
the topic of the next section. Here we conclude by highlighting the qualitative
explanation. In 
the chemical evolution literature terms, the L13 smoothly evolving model is equivalent to a system in which the metallicity
varies in response to a slowly changing $\Lambda(t)$. Among others, 
the behavior of $Z$ in the varying $\Lambda(t)$ case, has been also graphically and qualitatively discussed by Koeppen \& Edmunds (1999).
When, e.g., $\Lambda(t)$ slowly decreases with time,
the system evolves along the locus of the steady-state solutions on the $\mu -$Z plane, moving
towards lower gas fractions and higher metallicities. That is, 
from the steady state set by $\Lambda(t-dt)$ and $\eta(t-dt)$ to a new steady state (given by the current 
value of $\Lambda(t)$ and $\eta(t)$), where
the new $\mu_{\rm steady}$ is lower than the old one, whereas $Z_{\rm steady}$ increases (Koeppen \& Edmunds, 1999, their
Figs. 5 and 6).

\subsubsection{The accuracy of the L13 approximation: comparison to full
analytic solutions}

If one further adopts the initial condition $\mu=1$ (no stars), then 
$x_0=sSFR(0)/\nu$ diverges. Therefore we can assume $x_0=\infty$ in Eq.~\ref{appendix}.

In L13 the $sSFR$ decreases with time, driven by a decrease
in $\dot M_{acc}$. The variation is slow and at late times $|{d sSFR \over dt}|/sSFR << 1$; this
implies also a very small variation in the gas fraction ($\mu$ or equivalently $f$).
Therefore, there exists an epoch $t_1$ where $sSFR/\nu =x_1 << 1+\eta$ and
$\Lambda \simeq 1+\eta$. This implies that
${d\Lambda \over d x} \rightarrow 0$ when $t>t_1$ ($x < x_1$). 

As a consequence, we can write
$$\tilde I_2=\int_{x_0}^x {e^{\int_{\infty}^b F(a)da} \over \Lambda(b)^2} {d\Lambda \over d b} db \simeq
  \int_{\infty}^{x_1} {e^{\int_{\infty}^b F(a)da} \over \Lambda(b)^2} {d\Lambda \over d b} db \, ,$$
that is $\tilde I_2$ is effectively constant at $x < x_1$ (i.e. at late times).

At the same time, the exponent of the factor in front of $\tilde I_2$ grows in absolute value, therefore
the second term in Eq.~23 is:
$$e^{-const/x} \times \tilde I_2  \rightarrow 0$$
In this case Eq.~23 trivially reduces to:
\begin{equation}
Z\simeq {y_z \over \Lambda(x)}\equiv Z_{L13}\, .
\end{equation}

Similarly, $I_3$ is small because it has an exponentially declining factor and in the integral we have $|{\mathrm{d} sSFR/\nu \over \mathrm{d} t'}|<< (1+\eta+sSFR/\nu)^2$). We already discussed that $I_2$ has a fast exponential decline.
Therefore also the ``corrections'' given by $I_2$ and $I_3$ to the metallicity in Eq.~11 are small.

To reinforce our findings, in Fig.~\ref{fig_acc} we compare the numerical
integration of the metallicity (solid line) with these two limiting values (dashed - $Z_{L13}^{id}$, lower limit; dotted -
$Z_{L13}$, upper limit) for a particular set of $\nu$, $y_z$ and given
gas accretion history.
For the sake
of simplicity, we also arbitrarily set $y_z=0.02=Z_{sun}$, $\eta=0$ and $\nu =1$/Gyr.

The formula $Z=Z_{L13}^{id}={y_z \over 1+\eta+sSFR/\nu}$, used by L13 for the ideal regulator case (and $Z_A=0$)
gives always an excellent approximation, departing
from the numerical solution only by 0.1 dex at very late stages. It is the best approximations
of the true metallicity at the highest values of the sSFR.

On the other hand, the difference $Z-Z_{L13}$ is significant in the very early
phases of the evolution, when $\Lambda(t)$ and sSFR are un-correlated. This is however a consequence
of our set-up. In fact, assigning an initial $\mu=1$ leads the model
to evolve by consuming the gas
mass initially present in a way that is independent from the inflow, as a closed box.
A different initial set-up might reduce the difference between $Z_{L13}$ and the actual
metallicity in these early phases. At late times, instead, in this example, $Z_{L13}$
gives a very accurate approximation of the true metallicity.

\begin{figure}
\begin{center}
\includegraphics[width=3in]{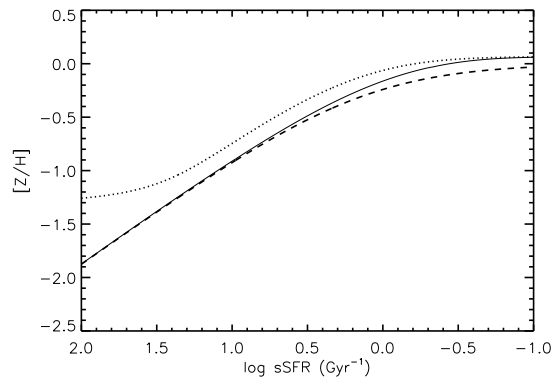}
\caption{Evolution of the metallicity as a function of the sSFR for galaxies evolving according
to the L13 model.
The solid line is the evolution given by the numerical integration of Eq.~\ref{eq:sol}, whereas the dashed
line is the metallicity approximated as $Z\sim Z_{L13}^{id}$ (Eq.~11, see text). Finally, the dotted line gives the metallicity
approximated by $Z_{L13}$ (Eq.~23).}
\label{fig_acc}
\end{center}
\end{figure}

\subsection{Special case III - The Closed Box Model in terms of the sSFR}

Before comparing the analytic approximate solutions to other numerical models,
we mention that, in the case of a model
with neither accretion nor outflows ($\Lambda=\eta=0$, Closed Box approximation, 
also known as the Simple Model),
the relation between metallicity and sSFR trivially is:
\begin{equation}
Z=y_z ln {1+sSFR/\nu \over sSFR/\nu} \,
\end{equation}
which, at early times (high sSFR), has the following approximate behavior:
\begin{equation}
Z\simeq  y_z \nu /sSFR \simeq  y_z \nu /(1+sSFR) \,
\end{equation}
which is very similar to $Z_{L13}^{id}$ when $Z_A=\eta=0$.
As the Closed Box Model well approximates the behavior of a model with $\Lambda(t)\ne0$
at early times (high gas fraction, e.g. Koeppen \& Edmunds 1999), 
this last equation is a good representation of the general
equation behavior in the regime of high sSFRs.

Therefore, we can conclude that
the sSFR, rather than $\Lambda$, seems to be the key quantity to accurately estimate the gas
metallicity of the system in a variety of cases.
The reasons lies in its close relation to the gas fraction $\mu$.

Also, in the Closed Box Model, the star formation has an exponential decline with
timescale $\tau=\nu (1-R)$. The results in this section then provide
a ready estimate for a self-consistent evolution of the metallicity and sSFR for
the widely adopted exponentially decaying star formation histories.

\begin{figure}
\begin{center}
\includegraphics[width=3.5in]{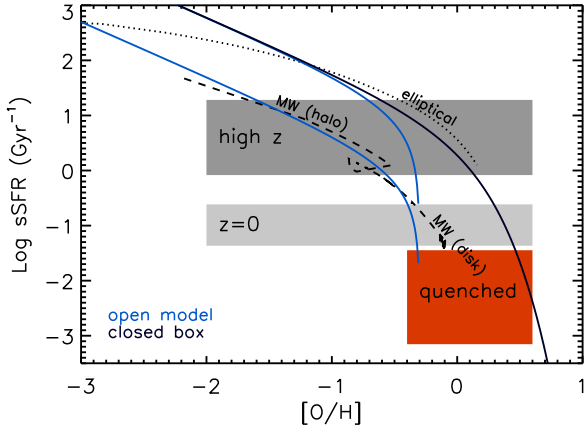}
\caption{Evolution of the metallicity as a function of the sSFR. %Models and symbols as in Fig.~\ref{fig2}.
The dashed line is the evolution predicted by a full numerical chemical evolution model
tuned to reproduce the Milky Way (MW) properties, whereas the dotted line
is the prediction for a model calibrated on both local ellipticals (Pipino \& Matteucci, 2004) and high redshift galaxies (Pipino
et al., 2011).
The solid lines give the metallicity approximated as $Z\sim Z_{L13}^{id}$ (see text) for models
with accretion (blue) and closed box (black). In this illustration, $Z_A=\eta=0$, $y_z=0.01$ and $\nu$ are matched to that used
in the full numerical chemical evolution models. Dark (light) grey areas give the typical values
of the sSFR observed in high-(low-)redshift galaxies, whereas the red box highlights the sSFR values for quenched
systems.}
\label{fig_acc_chem}
\end{center}
\end{figure}

\subsection{The accuracy of the L13 approximation: comparison to full
chemical evolution models}

The IRA is not a good approximation to
follow a system for a long ($>1$ Gyr) time, even if we focus on the total metallicity, which
is dominated by O (produced on a short timescale) and/or on metallicity inferred from O lines (among others).
That is, when $\mu <<$1 and after several Gyr of evolution, the effects of metals being recycled
by low mass stars cannot be ignored. We therefore further tested Eq.~\ref{sol1} against
the predictions of full numerical chemical evolution models calibrated on the abundance pattern
of the Milky Way and the properties of local ellipticals. We refer the reader
to Pipino \& Matteucci (2004), Pipino et al. (2011), and Calura et al. (2009) for a description of these models,
and to Pipino et al. (2013) for their predicted sSFRs, respectively.
The comparison is shown in Fig.~\ref{fig_acc_chem}, where the tracks in the metallicity-sSFR
plane are show for both analytical (solid) and numerical (dashed, Milky Way -MW; dotted, elliptical) models.
In the analytical models $\nu$ is matched to that used
in the full numerical chemical evolution simulations. For the same reason, we set $\eta=Z_A=0$.
The $Z\sim Z_{L13}^{id}$ approximation works well for a wide range in sSFR, becoming less accurate
at late times for the case of the Milky Way as expected. The difference is however less than
a factor of 2, comparable to the observational uncertainty in deriving the gas-phase metallicity.
As far as elliptical galaxies are concerned, we note that the closed box approximation (solid black line)
works better than the case with an infall with a long timescale (solid blue lines). This is not unexpected
since these galaxies should have formed on a short timescale (e.g., Matteucci, 1994), or equivalently, at high sSFR (e.g. Pipino et al., 2013 and references therein). 
In the numerical models, the galactic wind prevents 
the star formation to occur at arbitrarily low gas fractions (hence sSFR), whereas the ideal closed box
systems proceeds with $\mu\ \rightarrow\ 0$. To guide the eyes, dark (light) grey areas give the typical values
of the sSFR observed in high-(low-)redshift galaxies, whereas the red box highlights the sSFR values for quenched
systems at $z\sim 0$.

From this comparison, we can therefore conclude that a quasi-steady-state evolution depicted
in analytic models (as in L13) must be typical of
relatively low sSFR, disc galaxies, possibly representing the majority of the star forming ``main sequence'' at $z<2$.
For these galaxies, the current metallicity is well approximated by a steady-state-like value
set by the current sSFR.
We suggest here that ellipticals, instead, evolve at higher sSFR for a given metallicity than spiral galaxies for a given mass,
in a suggestive analogy to what happens in the [$\alpha$/Fe]-[Fe/H] plane (e.g. Fig. 4 in Matteucci \& Brocato, 1990). 
That is, highly $alpha$-enhanced stellar populations are a distinctive feature of galaxies
formed with high average sSFR, similar to those observed at $z>2$ (c.f. Pipino et al., 2013, see also discussions in Peng
et al., 2010, in the context of empirical models of galaxy growth, and in Pipino et al., 2009 -  their Sec. 3.2 - in the context of semi-analytical models of galaxy formation).
Clearly, having adopted the IRA to make the problem analytically tractable, any abundance ratio
predicted in the framework of this paper (and in L13) will be constant in time and simply equal to 
the ratio of the yields, unless one invokes selective inflows/outflows.
Therefore, we cannot predict an analytical quantitative relation between sSFR and $\alpha/Fe$ ratio
in the gas of a star forming galaxy.

We stress, however, that the ``morphology'' classes introduced in this section simply refer to the two typical parametrizations
adopted in numerical chemical evolution simulations. Namely high $\nu$ and quick infall are needed
to reproduce the chemical abundance pattern of present-day ellipticals, whereas smaller $\nu$ and longer
accretion histories seem to be typical of spirals, respectively. Therefore, in the context of this paper, such
``morphological'' classes should be understood as useful terms for linking the L13 model (and the more
general equations presented in this paper) to special cases of standard numerical models of chemical evolution.
A link between the actual morphology of the galaxies and the L13 model is beyond the scope of this paper.

\begin{figure}
\begin{center}
\includegraphics[width=3in]{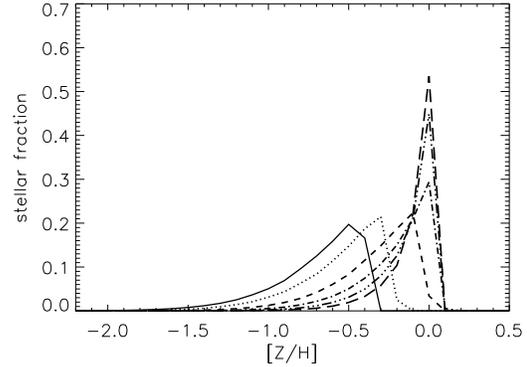}
\caption{Predicted stellar metallicity distribution for arbitrary galaxy models
evolving as prescribed by the L13 framework. See text for details.}
\label{fig5}
\end{center}
\end{figure}

Finally, one can exploit Fig. 2 as a diagnostic plot 
to readily estimate the star formation efficiency of a given galaxy observed at a given epoch,
by simply comparing its location in the sSFR-metallicity plane with a set of model predictions at fixed
yield (IMF) and varying $\nu$.

{ As mentioned above, the IRA does not hold at timescales comparable with those of massive stars. This would
have a noticeable effect if we were dealing with the chemical abundance ratios produced by a single stellar generation.
On the other hand, it is important to note that
the metal return of a stellar generation does not depend on the infall/outflow history.

In the earliest phases of the galaxy evolution, when the systems is not smoothly evolving,
 the biggest uncertainty in the derivation of the actual gas
metallicity Z is not related to the computation of the yield per se (and hence the assumption
of the IRA), but probably comes from the assumption of the ISM being always
homogeneous and well-mixed, as there might not be enough time for the metals to, e.g., cool down in new
star forming sites or to travel a long distance. 
Eq. 24 will still hold on a suitably chosen local level, if one replaces the infall metallicity $Z_A$ with that
inflowing from neighbouring regions, and considers the wind term will pollute the ISM immediately around
the star forming region. On a galaxy-wide scale, a suitable convolution of such a local version
applied to all star forming regions, would then give the overall metallicity evolution.}

\section{The metallicity in the stars of L13 galaxies}

\subsection{The stellar metallicity distribution}

Let us start by showing the expected stellar metallicity distribution (SMD)
in the framework of the L13 model, that is the fraction of stars per metallicity bin. In
order to illustrate the SMDs behavior we show in Fig.~\ref{fig5} a sample of galaxies smoothly evolving according
to L13. In this example, the galaxy final masses are in the range $10^{9-11} M_{\odot}$, 
and we assume that the more massive the galaxy, the higher $\nu$ (in the range 0.1-1.3/Gyr). We also assume that $Z_A=0$ and
a formation redshift, namely when the star formation rate is switched on, of $z_F=10$.
From the figure we can qualitatively infer an increase in the gas phase metallicity created by the variation in $\nu$ is mirrored by an increase
in the average stellar metallicity. In particular, the average metallicity in the most evolved (i.e., most massive) galaxies
already attained its uppermost boundary, set by the yield (e.g. Edmunds, 1990), $y_z=0.02$ in this illustration.
The larger star formation efficiency at high masses makes the SMD rather sharply peaked around $Z\sim y_z$.
Lower mass systems are still building up their SMD, which shows a long low-metallicity tail and a sharp cut-off
at $Z$ corresponding to the current gas metallicity.

\subsection{Gas versus stellar (average) metallicity}

Since the SMD of a galaxy is rarely accessible, it is useful to discuss other diagnostics that
involve, e.g., the mass-weighted stellar metallicity, defined as (Pagel \& Patchet 1975):
\begin{eqnarray}
Z_*={1\over M_*} \int_{M_0}^{M_*} Z(M) dM\simeq {y_z\over M_*} \int_{M0}^{M_*} {dM \over 1+\eta+sSFR/\nu}=\cr \cr \cr 
={y_z \nu (1-R)\over M_*} \int_0^{t} M(t') {sSFR(t')/\nu\over 1+\eta+sSFR(t')/\nu} dt'\, ,
\label{zstar}
\end{eqnarray}
where $M_*$ is the total mass of stars ever born contributing to the 
light at the present time, $Z(M)$ is the metallicity in the gas forming a given
stellar generation of mass $dM$, and we approximated the metallicity with $Z_{L13}^{id}$. 
For simplicity's sake, we neglect the metallicity of the infalling gas and assume $\eta=const$.

Next, we consider that
\begin{equation}
{M(t') sSFR(t') (1-R)} = SFR(t') (1-R)={d M\over dt'} \, .
\end{equation}
We make this substitution in Eq.~\ref{zstar}, and we then integrate the right-hand side by parts further
assuming that $M_0=M(0)\sim0$ (and hence
$sSFR(0)=\infty$). It follows that:
\begin{eqnarray}
Z_*\simeq{y_z \over M_*}
\times \Bigg({M_*\over 1+\eta+sSFR(t)/\nu}\; + \cr \cr
-\int_0^{t} M(t') {|dsSFR/dt'|\over \nu (1+\eta+sSFR(t')/\nu)^2} \, dt'\Bigg)\, ,
\label{zstarsol}
\end{eqnarray}
where we also consider that the sSFR decreases with time.
As for the gas metallicity solutions, it is easy to recognize a term in Eq.~\ref{zstarsol} that is similar to the L13
steady state/ideal regulator approximation for the current gas metallicity (Eq.~11 in this paper), and another containing the integral of the past history,
whose magnitude is related to the variation of the sSFR with time.

The gas-phase metallicity 
is an instantaneous measure, which should coincide
with the metallicity of the stars in the metal rich tail of the SMD, namely the most recently formed.
The average stellar metallicity
also accounts for the earlier (more metal poor) stellar generations, and hence it will be always lower
than the gas phase one, the difference being roughly given by the second term in the right-hand
side of Eq.~\ref{zstarsol}.

Generally, a large difference between gas and average stellar metallicity 
is found in the Closed Box Model, which features an exponentially decreasing
star formation history. Therefore most of the stars have a very low-metallicity (i.e., the classic
G-dwarf problem). The most extreme departure, however, is in the final stages, where $Z_{gas}\rightarrow\infty$ and $Z_*\rightarrow y_z$ for
$\mu\rightarrow 0$.
At the opposite end, a steady-state model ($\Lambda-\eta>1$, with both $\Lambda$ and $\eta$ constant in time)
has the property that $Z_*=Z_{gas}$ when it reaches equilibrium (see also Koeppen \& Edmunds, 1999), therefore
is this case we have the smallest difference at any time.

Models like the one discussed here (and in L13), have an intermediate behavior between these two cases. Qualitatively,
this can be understood as
these models tend to converge to
$\Lambda(t) -\eta \simeq 1$ at late times; therefore
both $Z_{gas}\rightarrow  y_z$ for small $\mu$  and $Z_* \le y_z$ (e.g., Edmunds, 1990), making  
the difference smaller than in the Closed Box Model case.
In earlier phases
of the evolution, the SFR is steadily increasing in time, making the SMD
strongly skewed toward large Z.
Therefore, at these early epochs, the youngest stellar generations (whose composition
is the same of the gas-phase) have a large
weight in the computation of the average stellar metallicity.
This finding implies that the evolution
in the stellar metallicity in the L13 framework can be approximated, for a ready and quick estimate, with a relatively simple dependence on the sSFR, that
mirrors that of the gas-phase metallicity (Eq.~\ref{appr1}).
Eq.~\ref{zstarsol} implies the existence of a mass-metallicity-SFR relation also in the case of the stellar metallicities, 
whereby galaxies evolving on tracks at higher sSFR have lower mass-weighted stellar metallicities. In the light of what has
been just discussed, this behavior should be detectable in the earliest phases, whereas it would become
less and less evident when the galaxy has passed the peak of its SMD, with $Z_*$ approaching the yield.

More quantitatively, in the L13 model we have $|dsSFR/dt|/sSFR \simeq 2.2/t$, namely decreasing with time and becoming less than 1
at times larger than $\sim 2.2$Gyr, that is roughly below $z=3$.
This means that, for most of the galactic evolution, the stellar metallicity is lower, but close, to the current
gas metallicity.

We illustrate this in Fig.~\ref{fig4diff}, where we show 
again the model galaxies presented in Fig.~\ref{fig5}: they feature an average stellar metallicity
lagging $<0.2$ dex behind the gas-phase metallicity at high redshift at any mass, where 
crosses, triangles and diamonds give the position on each track at $z=0$, $z=2$ and $z=3$, respectively.
The model predictions agree with the data (Halliday et al., 2008, Sommariva et al., 2012), which however feature large associated uncertainties.

\begin{figure}
\begin{center}
\includegraphics[width=3in]{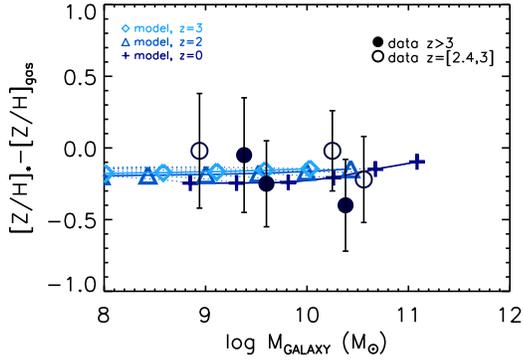}
\caption{Difference in metallicity between galactic components and its evolution as a function of the stellar mass. Models as in Fig.~\ref{fig5}.
Crosses, triangles and diamonds give the position on each track at $z=0$, $z=2$ and $z=3$, respectively.
Single galaxy measurements
in the redshift bins [2.4-3[ (empty circles) and [3,3.7] (full circles), respectively, as compiled by Sommariva
et al. (2012) are also shown.}
\label{fig4diff}
\end{center}
\end{figure}

\subsection{The mass- average stellar metallicity relation}

We note that
since the average metallicity in gas increases with galactic mass, we expect L13 model to predict also a stellar-mass
metallicity relation. We show that it is broadly 
consistent with observations in Fig.~\ref{fig4star}, where we plot the evolution in the stellar metallicity as a function of the 
stellar mass.
Crosses, triangles and diamonds give the position on each track at $z=0$, $z=2$ and $z=3$, respectively.
We also show the fit to the local observed relation (Panter et al., 2008, thick maroon line) and single galaxy measurements
in the redshift bins [2.4-3[ (empty circles) and [3,3.7] (full circles), respectively, as compiled by Sommariva
et al. (2012).

From the model-to-data comparison point of view, we highlight that, at $z\sim 0$, the predictions
for high mass galaxies seem to match the observations, whereas there is some tension at the low mass
end. On the theoretical side, the slope can be further steepened by acting on the relation between
$\nu$ and the initial mass.
The normalization of the predicted relation can be further adjusted by acting on the yield and of $z_F$. These
where however chosen to match the both the $z\sim 0$ and $z\sim 2$ mass-metallicity relation of the gas-phase.
On the observational side, a caveat is that $z\sim 0$ data include passive galaxies, which tend to be the most massive and metal
rich galaxies. We can therefore expect a milder observational slope in the mass-stellar metallicity relation
when selecting only star forming galaxies, in better agreement
with our model. The existence of metallicity gradients and aperture effects may further complicate the comparison
between data and models. { On the other hand, while we predict mass-weighted metallicities, the observables in questions
are luminosity-weighted quantities. The difference between luminosity-weighted and mass-weighted metallicity 
is negligible in massive, old, non-star forming galaxies
(e.g. Arimoto \& Yoshii, 1987), which make the high mass end of the local relation in Fig. 5. At smaller masses,
the mass-averaged Z are slightly larger than the luminosity-averaged ones, since the latter give more
weight to the earliest low-metallicity stellar populations (see e.g. Pipino et al., 2006). This may explain
the offset between z=0 predictions and observations at the low-mass end in Fig.5.}

In our models we do see an evolution in the metallicity with redshift at a given mass at $z<2$. This is
slightly ($\sim0.1$dex) smaller to that predicted in the gas phase, and shown in L13 (their Fig. 7). No apparent evolution 
is predicted between $z=2$ and $z=3$.
L13 model, however, predicts an evolution of the metallicity in this range, and the variation
can be readily estimate as follows: when the nebular emission line correction is included $sSFR(z\ge3)\sim6$/Gyr , whereas
$sSFR(z=2)\sim 2$/Gyr (e.g. Stark et al., 2013). If we apply Eq.~\ref{appr1},
we would infer an increase in metallicity
by a factor of $\sim 2$ (0.3 dex) between redshift $z\sim3$ and $z\ge2$ matching the observational data (e.g. Maiolino et al., 2008).
In the same time-frame, however, given these $sSFR$, galaxies increase their mass by at least a factor of 2, 
therefore they move almost diagonally in the $log Z_{gas}-log M_{GALAXY}$ plane. Since in these earlier phases
the average stellar metallicity tracks very well the gas metallicity, the model predicts a similar evolution
also in the $log Z_*-log M_{GALAXY}$ plane.
The combination of the two effects lead to an apparent non evolution in the mass-metallicity plane, as galaxies move along
a given track close to the 1:1 relation. They will then move up in metallicity at almost constant mass at $z<2$, when the sSFR is such that
the stellar mass increase is milder than the change in $Z$.

Sommariva et al. (2012), 
on the basis of the same data displayed in Fig.~\ref{fig4star}
seem to favor instead a lack of evolution at all redshifts.
It is however important to note the large observational errors and the scatter
affecting the $z>2$ measurements of a few single galaxies. 

For comparison, we display the evolution of a galaxy evolving as a closed box model (solid black line) in Fig.~\ref{fig4star}.  We 
can safely conclude that the data strongly disfavor closed box and favor flow-through models like that of L13.

It also useful to remind the
danger of { comparing mass-weighted predicted quantities to luminosity-weighted observables}. 
This effect is more important at high z, when galaxies feature high SFRs. We stress
again that the $z\sim 0$ data mix active and passive galaxies, whereas the high redshift data-points refer
to star-forming galaxies. Moreover, while $z\sim 0$ metallicities are mostly related to the optical part
of the spectrum, in high redshift galaxies are derived by means of UV absorption lines (Rix et al., 2004). 
Therefore a direct and robust, entirely empirical, comparison
between the metallicity of the bulk of the stars in galaxies at different redshift has yet to come.

\begin{figure}
\begin{center}
\includegraphics[width=3in]{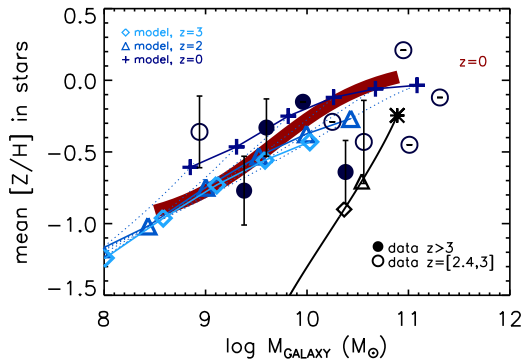}
\caption{Stellar metallicity evolution as a function of the stellar mass. Models as in Figs.~\ref{fig5} and~\ref{fig4diff}.
Crosses, triangles and diamonds give the position on each track at $z=0$, $z=2$ and $z=3$, respectively.
The predictions for the Closed Box Model system with $\nu=0.3$/Gyr are shown by the (almost vertical) solid black line.
The fit to the local observed relation (Panter et al., 2008, thick solid line) and single galaxy measurements
in the redshift bins [2.4-3[ (empty circles) and [3,3.7] (full circles), respectively, as compiled by Sommariva
et al. (2012) are also shown.
}
\label{fig4star}
\end{center}
\end{figure}

%\section{Discussion}

\section{Discussion and Conclusions}

\subsection{The gas phase metallicity}

\subsubsection{On the variation of the $Z$ as a function of the sSFR evolution}

In this paper we explored the dependence of gas-phase metallicity
on the specific star formation rate. 
In particular, we derived general analytic formulae
that relate the gas phase metallicity to both the infall-to-star formation rate ratio and
the sSFR, for the case of single-zone single-phase galaxies
and a linear Schmidt (1959) star formation law.
The derived relations take the typical form {$Z(sSFR)={y_Z \over \Lambda(sSFR/\nu)}+I(sSFR/\nu)$, where $I$ is the integral of the past enrichment history over the sSFR.}

In this article, both the inflow- and the outflow-to-star formation rate ratios are functions of time (equivalently of the sSFR)
and may depend on the input cosmology, the amount of gas that may penetrate the star forming regions
of galaxies, and the adopted star formation law.
It is important to stress that this approach is different from that adopted
in many analytical chemical evolution works in the literature, and still in 
use to interpret the metallicity of galaxies. These studies adopt $\Lambda$ and $\eta$ constant, that is they
do not take into account that realistic outflow- and inflow-to-star formation rate ratios may change
with time as the result of the evolution of the galaxy. Therefore, when they are compared to data at a given redshift,
they can only give a simple parameterized
understanding of the mass-metallicity relation at that fixed epoch, rather than offering a comprehensive
view of galaxy evolution with time.

We show that \emph{in many circumstances (early evolution, quasi-steady state evolution with slowly decreasing
sSFR) a good estimate of the gas-metallicity is obtained by the value $Z_{L13}^{id}$
that the system would have if in steady-state evolution with the infall-to-star
formation ratio set by the current value of the sSFR} (Eq.~11), that
is as in the ideal regulator case of L13.
On the other hand, a metallicity obtained by current value of the infall-to-star formation
rate, i.e. $Z_{L13} $(Eq.~8, L13 non ideal regulator case), 
would slightly overestimate the current metallicity of the system.
These two values bracket with high accuracy the current metallicity of the system.
Therefore, we provide the formal justification to L13 approximations and extend their validity to a larger range of cases.
In particular, \emph{the formula adopted by L13 for the ideal case is exact for systems with $\Lambda-\eta>1$ and $\Lambda , \eta = const$.
It is also a good approximation of the actual metallicity for galaxies with slowly
decreasing accretion rates}.

Also, we add that, since we did not specify anything on both the size and the geometry of what we called the ``galaxy'', the equations
in principle hold at both the \emph{local} (i.e. for each star forming region) and the \emph{global} galaxy level,
with the latter being a suitable weighted average of the single star forming regions. Such a theoretical
expectation seems to be corroborated by very recent observations (Rosales-Ortega et al., 2012).

Finally, L13 (their Fig. 7) show the predictions for the mass-metallicity relation at $z>0$ in some specific cases
calibrated on either the Mannucci et al. (2010) or the Tremonti et al. (2004) $z\sim0$ relations.
A qualitative agreement with the $z\sim 2$ data is achieved.
In this paper we do not repeat the exercise. However, { in Fig.~6, we plot the $z<4$ tracks of the models
shown in Fig.~5, compared to the full three-dimensional fundamental metallicity relation as given
by Mannucci et al. (2010, their Eq.~2). In the L13 framework, galaxies evolve along the surface given
by the fundamental metallicity relation.}

Also, we wish to highlight the following point, which 
was not discussed in L13, but it is implied by the assumed $Z=f(sSFR)$ relation at a fixed epoch.
It is in fact relevant to the data-model comparison to note that the galaxies in the $z=2$ and $z=3$ observational samples
show a rather flat SFR-mass relation (c.f. Mannucci et al., 2009, their Fig. 6), by virtue
of their selection. This relation is flatter than the typical SFR-mass relation for star forming galaxies at
the same epoch (e.g. Daddi et al., 2007). If we assume that the observational samples are culled out
from the star forming population at their respective redshifts, the selection of the most massive galaxies
with systematically below the average SFR creates
a bias such that the most massive galaxies tend to systematically have the lowest sSFR, and thus to be
the most metal rich at their mass scale (at least this is the expectation in the L13 theoretical framework). By virtue
of the SFR selection threshold, the low mass galaxies will have higher than average SFR, and hence a lower
metallicity than the typical star forming galaxy at the same redshift and mass.
This means that the observational samples might be biased in the direction of having
a steeper mass-metallicity relation than the typical relation of an unbiased sample of star forming galaxies
at the same redshift. Therefore any empirical conclusion on the evolution in the slope of the mass-metallicity
relation must be treated with caution.

\subsubsection{On the L13 metallicity formula: accuracy and comparison to full numerical chemical evolution models }

In order to quantify the accuracy of the L13 approximations, we compared them to a full and
direct numerical integration of the same equations, finding an \emph{excellent agreement ($<0.1$ dex) for three orders of magnitude in the sSFR and almost
2 dex in $Z$}.

By comparing tracks in the $Z-sSFR$ plane given by either the L13 approximation or the Closed
Box relation to the predictions of full numerical chemical evolution
models which relax some of the simplifying assumptions adopted in the analytic case, we find that
\emph{star forming $z<2$ (spiral) galaxies, where the sSFR slowly decreases with time,
the system evolves along the locus of the steady-state solutions of decreasing
gas fraction and increasing metallicity, exactly as in the L13 gas-regulated model}.
In particular,
these system seeks the steady state metallicity without attaining it and the current metallicity is set by the current value of the sSFR.

\emph{Fast-forming (elliptical) galaxies, evolve at higher sSFR than slowly-evolving systems at the same
metallicity, with a remarkable similarity to the well-known behavior in the [$\alpha$/Fe]-mass plane}.
Their track in the sSFR-Z plane is better approximated by Closed Box models.\\

\subsubsection{The SFR as the second parameter in L13 and other special cases (closed
box, evolution at constant gas mass)}

The { actual functional form} of a mass-(s)SFR-metallicity relation is quite controversial,
with empirical findings also including claims of a reversal (namely high SFR would correspond to high metallicity) 
at high stellar masses (e.g. Yates et al., 2012) and a lack of any SFR effects at all masses (e.g. Sanchez et al., 2013).
Clearly, differences may originate from a variety of empirical issues related to the sample specifics (including
redshift range and aperture effects, e.g. Sanchez et al., 2013)
as well as to the methods used to derive the metallicity (and the SFR).

In L13 the metallicity dependes inversely on the sSFR.
To some extent we expect a smaller dependence on the sSFR as a second parameter at high masses,
simply because in L13 $\nu$ becomes larger and hence the term $sSFR/\nu$ smaller than the other terms
in the denominator of Eq. (8). In other words,
the most massive models settle earlier on an evolutionary track where the 
metallicity quickly asymptotes to the yield, and the second parameter effect caused by variations in the (s)SFR
become consequently small.
It seems more difficult to explain a reversal of the trend above a given stellar mass scale. 

Moreover, L13 model is meant to reproduce the average galaxy, therefore, it 
does not take into account that episodic bursts and mergers may also happen and move galaxies further out
of the ``average'' quasi steady-state evolution represented by our tracks.\\

As a matter of fact, in this paper \emph{we also show that 
an anti-correlation between $Z$ and sSFR is found also in the early evolutionary phases of the Closed Box model}.

In other works (e.g. Dav\'e et al., 2012), the case $\Lambda-\eta=1$ (which is a generalization
of Larson 1972's \emph{extreme infall}
in the context of analytic chemical evolution)
has been dubbed ``steady-state''. In other words, all the net accreted gas is used up to form
stars. More specifically, it is
the $\Lambda=1,\eta=0$ case which is known as the ``extreme infall''. It has the property
of preserving the gas mass, rather than the gas fraction, and that the metallicity would evolve as 
$Z=(Z_A+y_z) \, (1-exp(1/\mu -1))$, asymptotically approaching the yield for $\mu$ approaching 0.

The generalization of \emph{extreme infall} where both inflows and outflows are present ($\Lambda$, $\eta=\, const$), has
the following analytical solution for the metallicity:

\begin{equation}
Z={(Z_A \Lambda +y_z)  \over \Lambda}\biggl\lbrace 
1 - e^{- \Lambda \, (1/\mu -1)}\biggr\rbrace 
\end{equation}
or equivalently:

\begin{equation}
Z={[Z_A (1+\eta) +y_z]  \over (1+\eta)}\biggl\lbrace 
1 - e^{- (1+\eta) \, (1/\mu -1)}\biggr\rbrace
\label{eq:ei}
\end{equation}

It is important to note here that, despite assuming the validity of the condition $\Lambda-\eta\simeq1$, Dav\'e et al. (2012) do not
fully derive these solutions. In fact, they base their model on their Eq. 9. We find that their formula can be re-arranged, after discarding
the trivial solution $Z(t)=0$ and assuming that $Z\ne Z_A$, as: $Z\sim {[Z_A (1+\eta) +y_z]  \over (1+\eta)}$,
which is only an approximation to our exact solutions (e.g., Eq.~\ref{eq:ei}), valid when
the gas fraction is small (as pointed out also by Dayal et al., 2013). This latter condition ($\mu << 1$) is unlikely to be true in high redshift galaxies.

When the galaxy is in its asymptotic regime at constant gas mass, that is $Z\simeq Z_A+y_z$, the only way to increase its metallicity is by acting
on $Z_A\ne 0$. In the first place, in the light of our full analytic derivation, we stress that the correct solutions for $Z$ in a  standard analytic chemical evolution model when the infalling gas metallicity changes with time and it is linked to the past history of the galaxy,
must take into account that $Z_A=Z_A(t)$ in the formal integration (Eq.~22 in this paper, see also the implementation of  galactic fountains in Recchi et al., 2008).

We then note that when $Z_A$ drives the metallicity, it
increases with time in a manner that is not necessarily linked to the sSFR evolution.
In other words, in systems with $\Lambda(-\eta)\sim1$
the gas fraction still changes with time, leading to changes in the sSFR which are un-correlated
to variations in the gas metallicity (in principle set by yield, and varied through a changing metallicity in the
``re-accretion'' of previously ejected material).
This also implies that the scatter around the average $Z-sSFR$ relation cannot be described by the same equation
that governs the $Z=Z(sSFR)$ evolution as in L13. On the contrary, in Dav\'e et al. (2012), the explanation of the scatter (and
of the $Z-sSFR$ anti-correlation)
requires stochastic events that drive the galaxies out of equilibrium, either enhancing the SFR (e.g. mergers)
or momentarily suppressing it (e.g. a sudden decrease in the accretion), and causing either a decrease
or a increase in $Z$, respectively. This perspective is not dissimilar to the explanation
given by Mannucci et al. (2010) when they first presented the empirical results on the ``fundamental
metallicity relation'', and it is further extended in other recent work (e.g. Forbes et al., 2013) which depict
both the mass-SFR and the mass-Z relations as the result of statistical equilibrium in the galaxy population
at a given epoch.

\begin{figure}
\begin{center}
\includegraphics[width=3in]{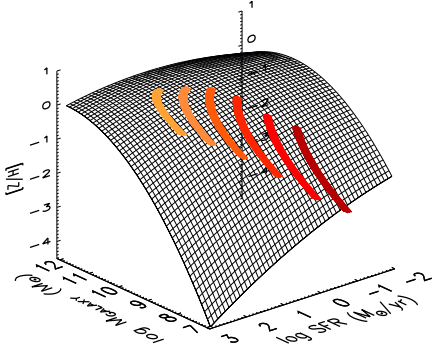}
\caption{$z<4$ evolution of the models discussed in Figs.~\ref{fig5} and~\ref{fig4diff} in the
three dimensional space given by star formation rate, mass and metallicity. The shaded area
is the analytical formula of the fundamental metallicity relation given by Mannucci et al. (2010).
}
\label{fig3d}
\end{center}
\end{figure}

\subsection{Stellar metallicities in the L13 model}

As a further extension of the L13 model, we show that \emph{it
also naturally predicts a mass-metallicity relation in the stellar component which
matches the current data at different epochs}. 

We compare our predictions to the data, and despite the encouraging qualitative agreement,
no firm quantitative conclusions can be drawn due to: i) the large
scatter in the high redshift data; ii) the lack of consistency among the stellar metallicity measurements
at different epochs; and iii) the presence of both passive and star forming galaxies in the $z\sim0$ dataset.

L13's slowly evolving galaxies therefore match both the observed cosmic evolution in the gas and in the average stellar
metallicity at a given mass. This is a consequence of the fact that
the \emph{average stellar metallicity systematically lags $\sim 0.1-0.2$ behind the gas metallicity of the same galaxy}
(as observed).

In evolved ($\mu << 1$) systems, such a small difference can be explained by the fact
the both the gas and the average stellar metallicity tends to the yield. Whereas, during earlier stages
of the evolution, the explanation lies in the fact that the SFR is steadily increasing in time. Therefore the youngest stellar generations (whose composition
is the same of the gas-phase) have a larger
weight in the computation of the average stellar metallicity.

These findings imply that the evolution of the average stellar metallicity in the early phases of L13 galaxies as
a function of the sSFR can be approximated by the same formula adopted for the gas phase metallicity.

%\end{itemize}

\section*{Acknowledgments}
We thank the referee for the comments that improved the quality of the paper.

\section*{Appendix}

\subsection*{The impact of the assumed star formation law.}
{ 
Let us discuss the impact of a more general star formation law on our results, that, for the purpose of this discussion,
we write as $\psi=\nu M_{gas}^{(1+x)}$. In this paper we presented
the results for the case $x=0$.
Such a linear Schmidt volumetric relation is a standard assumption in the literature and it has the advantage
of a slight simplification of the calcultions presented in this paper. To corroborate our assumption, we note that
in a recent paper, Krumholz et al. (2012) showed that a simple volumetric star formation law as the one adopted in our paper, can explain a
wide range of both local and high-redshift observations.

Furthermore, it leads to a relation with exponent $1.5$ if
the star formation efficiency is expressed in units of the local free fall time, and this latter quantity is in turn expressed
as a function of the gas volume density. This also ensures compatibility with the expression adopted in studies where SFR and density
are in units of surface which assume an exponent $x\sim 1.4$.

It is well known that the solutions of the form $Z=Z(\mu)$ of analytical chemical evolution models do not explictly depend on the 
SFR (and its law). Therefore, the particular star formation law adopted does not influence these general results. 
It is the conversion of the gas fraction into sSFR that introduces a dependence on the assumed star formation law in the equations
of the form $Z=Z(sSFR)$. To see the impact of the change let us proceed
as in the main body of the paper, namely let us focus on the steady-state solutions and the derive more general statements. 

In the case of the steady state, the results presented in Eqs. 12, 13, 14 and 16 (its first row),
as well as other results like Eqs.~32 and 33, will not depend on $x$ as they do not feature any
explicit dependence on the star formation law.

On the other hand, when $x\ne 0$,
Eq. 1 would be $$sSFR/\nu \propto {\mu \over 1+ \mu} \mu^x$$ (e.g. Reddy et al. (2006), when x=0.4), and Eq. 15 would read 
as $$sSFR/\nu=(\Lambda-\eta-1) M_{gas}^x \, .$$
The steady state solution presented in Sec. 3.2 (Eq. 16 second row)  will then be 
$$Z={1 \over 1+\eta+sSFR/(\nu M_{gas}^x)}\, ,$$ where
for simplicity we ignore the metallicity of the infalling gas. 
If $x>0$, the system behaves as if it has a higher effective star formation efficiency $\nu M_{gas}^x$.
Moreover $x\ne 0$ would imply an evolution
at constant gas fraction and metallicity with the sSFR still changing in time as $M_{gas}^{x}$.

Given the close link between $I_1$ (Eq.11), Eq. 24 and the steady-state solution (Eq. 16), we expect a similar variation, 
namely the apperarence of a factor $\sim M_{gas}^x$ 
in the expression for $Z_{L13}$ in both the ideal and non indeal case,
as well as in the general solutions. More quantitatively, this happens because
the term $G$ in the differential equation 19 will now be:
\begin{equation}
G(sSFR)={1+{Z_A \Lambda \over y_z} \over sSFR/\nu ((1+x)(\Lambda-\eta-1)M_{gas}^x-sSFR/\nu)} \, ,
\end{equation}
leading to a change in the expression for $F(sSFR)$ too.
The qualitative description of the galaxy behavior will not change: as
these models tend to a $\Lambda\sim 1$ (constant gas mass) evolution in the long term, 
the factor $M_{gas}^x$ will be merely a constant for all practical purposes.}

%.................................................................................................................

\end{document}